\documentclass[aps,prl,a4paper,twocolumn,longbibliography,superscriptaddress,nofootinbib]{revtex4-1}
\usepackage{amsmath}
\usepackage{amssymb}
\usepackage{mathtools}
\usepackage{graphics}
\usepackage{xcolor}
\usepackage{comment}

\newcommand{\beq}{\begin{equation}}
\newcommand{\eeq}{\end{equation}}

\newcommand{\bk}{{\mathbf k}}
\newcommand{\wc}{\tilde{\omega}}

\newcommand{\bp}{{\mathbf p}}

\newcommand{\cT}{{\mathcal T}}
\newcommand{\ctT}{{\tilde{\cT}}}
\newcommand{\omegazero}{\omega^{(0)}}

\usepackage{hyperref}
\hypersetup{%
   pdfpagemode=None, 
   pdfstartpage=1,
   pdfmenubar=true, pdftoolbar=true,
   colorlinks = true, linkcolor=blue, citecolor=blue, urlcolor=blue,
   bookmarksopen=false
 }

\begin{document}
\title{Bose polarons at finite temperature and strong coupling}

\author{Nils-Eric Guenther}
\affiliation{ICFO -- Institut de Ciencies Fotoniques, The Barcelona Institute of Science and Technology, 08860 Castelldefels (Barcelona), Spain}
\author{Pietro Massignan}
\email{pietro.massignan@upc.edu}
\affiliation{Departament de F\'isica, Universitat Polit\`ecnica de Catalunya, Campus Nord B4-B5, E-08034 Barcelona, Spain}
\affiliation{ICFO -- Institut de Ciencies Fotoniques, The Barcelona Institute of Science and Technology, 08860 Castelldefels (Barcelona), Spain}
\author{Maciej Lewenstein}
\affiliation{ICFO -- Institut de Ciencies Fotoniques, The Barcelona Institute of Science and Technology, 08860 Castelldefels (Barcelona), Spain}
\affiliation{ICREA, Pg.\ Lluis Companys 23, 08010 Barcelona, Spain}
\author{Georg M.\ Bruun}
\affiliation{Department of Physics and Astronomy,  Aarhus University, Ny Munkegade, DK-8000 Aarhus C, Denmark}

\begin{abstract}
A mobile impurity coupled to a weakly-interacting Bose gas, a Bose polaron, displays several interesting effects. While a single attractive quasiparticle is known to exist at zero temperature, we show here that the spectrum splits into two quasiparticles at finite temperatures for sufficiently strong impurity-boson interaction. The ground state quasiparticle has minimum energy at $T_c$, the critical temperature for Bose-Einstein condensation, and it becomes overdamped when  $T\gg T_c$. 
The quasiparticle with higher energy  instead exists only below $T_c$, since it is a strong mixture of the impurity with thermally excited collective Bogoliubov modes. This phenomenology is not restricted to ultracold gases, but should occur whenever a mobile impurity is coupled to a medium featuring a gapless bosonic mode with a large population 
for finite temperature.
\end{abstract}

\maketitle

Mobile impurities in a quantum bath play a fundamental role in a wide range of systems including metals and dielectric materials~\cite{Mahan2000book}, semiconductors~\cite{Gershenson2006}, $^3$He-$^4$He mixtures~\cite{BaymPethick1991book}, and high-$T_c$ superconductors~\cite{Dagotto1994}. In certain limits, they provide a paradigmatic realization of Landau's fundamental concept of a quasiparticle. 
The precise experimental measurements on impurity atoms in ultracold Fermi gases~\cite{Schirotzek2009,Kohstall2012,Koschorreck2012,Cetina2016,Scazza2017} combined with several theoretical investigations~\cite{Chevy2006,Prokofev2008,Mora2009,Punk2009,Combescot2009,Cui2010,Massignan2011,Massignan2014,Wei2015} led to fundamental insights into this problem.
Recently, two experimental groups embedded impurity atoms in a Bose-Einstein condensate (BEC) and observed long-lived quasiparticles coined Bose polarons~\cite{Jorgensen2016,Hu2016}.  
Bose polarons have been investigated  using a variety of theoretical techniques~\cite{Astrakharchik2004,Cucchietti2006,Li2014,Levinsen2015,Shchadilova2016,Tempere2009,Rath2013,Christensen2015,Grusdt2017,Schmidt2016,Ardila2015,Ardila2016}, 
and they have also been considered in one dimension~\cite{Catani2012,Volosniev2015,Grusdt2017bis,Lampo2017,Volosniev2017}.

A qualitatively new feature of the Bose polaron with respect to polarons in a Fermi gas or in solid state systems is that the environment undergoes a phase transition to a BEC at $T_c$. 
This changes drastically the low-energy density-of-states  of the environment, and should therefore affect significantly the  polaron. 
Temperature effects on Bose polarons have been examined so far only theoretically, either in the mean-field regime~\cite{Boudjemaa2014},  at high  temperature \cite{Sun2017}, or for immobile Rydberg atoms~\cite{Schmidt2016}.  

Here, we  develop a strong coupling diagrammatic scheme designed  to include  scattering processes important for finite temperature.
Using this, we show that the polaron splits into two quasiparticle states for $0<T<T_c$ for strong attractive coupling. 
The energy of the lower polaron depends non-monotonically on temperature with a minimum at $T_c$, whereas the energy of the upper polaron  
increases until its quasiparticle residue vanishes at $T_c$.
The generic mechanism causing these effects is the coupling between the impurity and low energy Bogoliubov modes with an infrared divergent population for finite $T$. 
Consequently, similar effects should occur whenever a mobile impurity is coupled to a gapless bosonic mode, as for example in Helium liquids or quantum magnets. Indeed, an 
analogous  splitting has been predicted for quasiparticle modes in hot electron and quark-gluon plasmas \cite{Klimov1982,Weldon1982,Weldon1989,Baym1992}, providing an interesting link between low and high energy quantum phenomena.

\begin{figure}[t]
\includegraphics[width=\columnwidth]{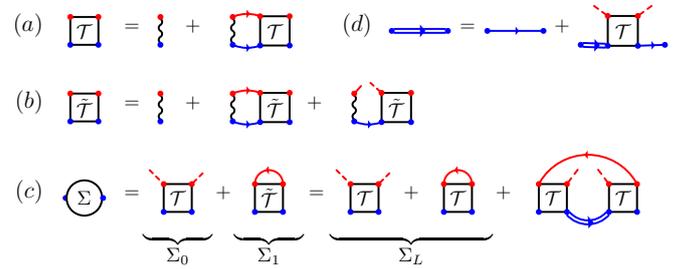}
\caption{\label{fig:Model}
Feynman diagrams yielding the self-energy $\Sigma$ within the ``extended" ladder approximation. 
Thin blue lines represent the bare impurity propagator, 
solid red lines are Bogoliubov propagators, 
dashed red lines are condensate bosons,
and wavy lines represent the vacuum scattering matrix $\cT_v$. 
The ladder self-energy is indicated by $\Sigma_L$, and the double blue line is an impurity dressed by the condensate only.
}
\end{figure}

\paragraph{System.--}
We consider a single impurity particle of mass $m$, immersed in a gas of weakly-interacting bosons of mass $m_B$ and density $n$ at temperature $T$.
We take $\hbar=k_B=1$, and  measure momenta and energies in units of $k_n = (6\pi^2 n)^{1/3}$ and $E_n = k_n^2/2 m_B$.  
The boson-boson interaction is short-ranged and characterized by a scattering length $a_B$, and the condition of weak interaction means $0<k_na_B \ll 1$. 
Below the critical temperature $T_c=[4/(3\sqrt\pi \zeta(3/2))]^{2/3}E_n\approx0.436E_n$, the bosons are accurately described using  Bogoliubov theory~\cite{Fetter1971}.
The condensate density is $n_0=n[1-(T/T_c)^{3/2}]$ and the chemical potential is $\mu_B=\cT_B n_0$ with  $\cT_B = 4\pi a_B/m_B$.
The dispersion of the excitations reads $E_{\mathbf k}=[\epsilon^B_{\mathbf k}(\epsilon^B_{\mathbf k}+2\mu_B)]^{1/2} $,
with  $\epsilon^B_{\mathbf k}=k^2/2m_B$. 
Above $T_c$ the condensate disappears, the excitations become free bosons, $E_{\mathbf k}=\epsilon^B_{\mathbf k}$, and the chemical potential reduces to that of an ideal Bose gas. 
The impurity interacts with the bath through a short-ranged potential described by the $s$-wave scattering length $a$.
Given that we study a single impurity, we consider its effect on the Bose cloud negligible.

\paragraph{Diagrammatic analysis.--}
An impurity with momentum $\mathbf k$ is described  by the  imaginary-time  Green's function 
$\mathcal{G}({\mathbf k},i\omega_j)=1/[\mathcal{G}_0({\mathbf k},i\omega_j)^{-1}-\Sigma({\mathbf k},i\omega_j)$], 
where $\mathcal{G}_0({\mathbf k},i\omega_j)=(i\omega_j-\epsilon_{\mathbf k})^{-1}$  with $\epsilon_{\mathbf k}=k^2/2m$ is the Green's function of a free impurity, and $\omega_j=2\pi jT$ is a Matsubara frequency~\cite{Fetter1971,footnote}. 
The key quantity to calculate is the impurity self-energy $\Sigma({\mathbf k},i\omega_j)$, and in presence of strong interactions we must resort to suitable approximations. 
As a minimum, we have to include the ladder diagrams of Fig.\ \ref{fig:Model}(a), which describe the energy shift due to two-body processes between the impurity and Bogoliubov excitations.
This yields the ladder scattering matrix $\cT(\bp,i\omega_j) = 1/\left[\cT_v^{-1}-\Pi(\bp,i\omega_j)\right]$, where $\cT_v=2\pi a/m_r$ is the vacuum scattering amplitude, $m_r=m m_B/(m + m_B)$ is the reduced mass, and
$
\Pi(\bp,i\omega_j) =
\int \! \frac{d^3k}{(2\pi)^3} \!
\left(
\frac{ u_{\mathbf k}^2(1+f_\bk)}{i\omega_j - E_{\mathbf k}-\epsilon_{{\mathbf k}+{\mathbf p}}}
+\frac{ v_{\mathbf k}^2 f_\bk} {i\omega_j + E_{\mathbf k}-\epsilon_{{\mathbf k}+{\mathbf p}}} + \frac{2 m_r}{k^{2}}
\right)
$
is the renormalized pair propagator for an impurity and a boson from the bath.
Here $f_\bk=1/[\exp(E_\bk/T)-1]$ is the Bose distribution, 
and $u_{\mathbf k}^2=[(\epsilon^B_{\mathbf k}+\mu_B)/E_{\mathbf k}+1]/2$ and 
$v_{\mathbf k}^2=u_{\mathbf k}^2-1$
are the Bogoliubov coherence factors.

A very recent perturbative analysis~\cite{Levinsen2017} showed that events where the impurity scatters excited bosons into the BEC are important  at finite temperatures.
To include these events, which are not contained in the ladder approximation, we introduce the ``extended" scattering matrix $\ctT $ shown in Fig.~\ref{fig:Model}(b), and given by 
\beq\label{extTmatrix}
\ctT(\bp,i\omega_j) = \frac{1}{\cT_v^{-1}-\Pi(\bp,i\omega_j)-n_0\mathcal{G}_0(\bp,i\omega_j)}.
\eeq 
Compared to the ladder approximation $\cT$, the denominator of $\ctT$ contains the additional term $n_0\mathcal{G}_0(\bp,i\omega_j)$, which describes pair 
propagation where the entire momentum is carried by the impurity while the boson is in the BEC. This process is represented by the third term in Fig.~\ref{fig:Model}(b). 

As shown in Fig.~\ref{fig:Model}(c), within the extended scheme the impurity self-energy is $\Sigma=\Sigma_0 + \Sigma_1$, where  
\begin{align}
\Sigma_0(\bp,i\omega_j) = n_0 \cT(\bp,i\omega_j)
\end{align}
is the energy shift experienced by the impurity through interactions with the condensate only, and 
\begin{multline}\label{Sigma1}
\Sigma_1(\bp,i\omega_j)=
\int \frac{d^3k}{(2 \pi )^3}  \Big[ u_\bk^2 f_\bk \ctT(\bk+\bp,i\omega_j+E_\bk) \\
+ v_\bk^2 (1+f_\bk) \ctT(\bk+\bp,i\omega_j-E_\bk)\Big]
\end{multline}
is the energy shift coming from interactions with bosons excited out of the BEC. Note that replacing $\cT$ with $\ctT$ in $\Sigma_0$ is not allowed, since this would lead to double counting of diagrams. As shown in  Fig.~\ref{fig:Model}(c), $\Sigma_1$ may be decomposed in two contributions: one where the impurity scatters a boson from an excited state to another (second diagram), and one where the excited bosons are scattered virtually back into the BEC (last diagram).
An approximation similar to ours was used in Ref.~\cite{Guidini2015} for analyzing Bose-Fermi mixtures, while  the ladder approximation $\Sigma_L$ (considered at $T=0$ in Ref.~\cite{Rath2013}) is recovered by replacing $\ctT$ with $\mathcal T$ in (\ref{Sigma1}), which is equivalent to neglecting the rightmost  diagram in Fig.~\ref{fig:Model}(c).

The spectral function of the impurity particle with momentum $\bp$ is given by $A(\bp,\omega)=-2\textrm{Im}[G(\bp,\omega)]$, 
where we have performed the usual analytic continuation $i\omega_j\rightarrow \omega+i0_+$, suppressing $i0_+$ for notational simplicity. A quasiparticle corresponds to a sharp peak in the spectral function. Its energy is found by solving 
\begin{align}
\omega_\bp=\epsilon_\bp+\text{Re}[\Sigma(\bp,\omega_\bp)],
\label{QPeqn}
\end{align}
and the quasi-particle is well-defined when the damping $\Gamma=-Z_\bp\text{Im}[\Sigma(\bp,\omega_\bp)]$  is small. Here
$Z_\bp=1/\{1-\partial_\omega\text{Re}[\Sigma(\bp,\omega)]|_{\omega_\bp} \}$
is the residue, a measure for the spectral weight of the quasiparticle peak. 

\paragraph{Results.--}
We now show numerical results for the properties of the dressed impurity, taking for concreteness  equal impurity and boson masses $m=m_B$  corresponding to the  Aarhus experiment~\cite{Jorgensen2016}.  Having this experiment in mind, we also focus on zero momentum polarons. Unless otherwise specified, we set $k_n a_B = 0.01$.

\begin{figure}[t]
\includegraphics[width=\columnwidth]{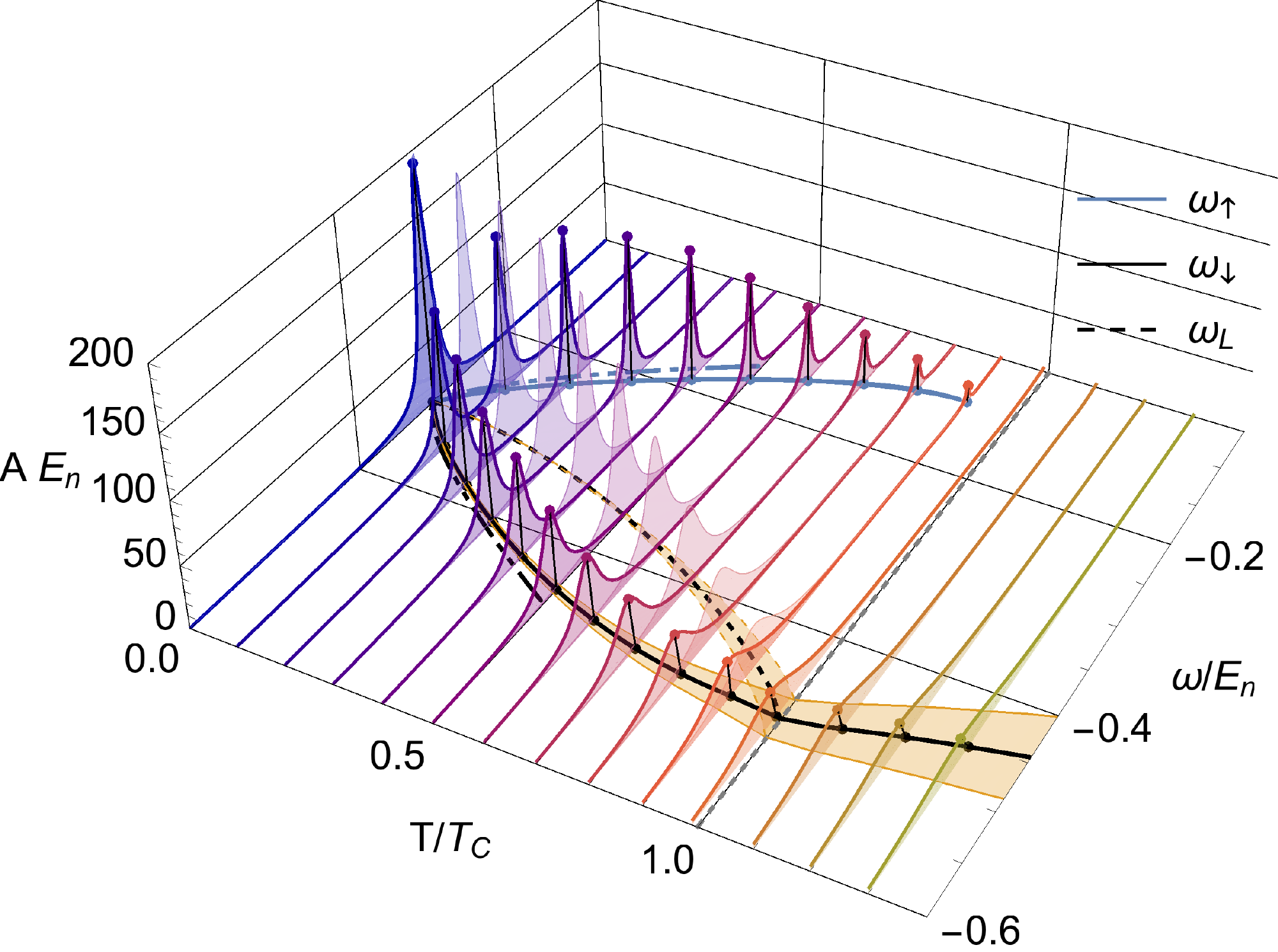}
\caption{\label{fig:attractivePolaronsKnAM1}
Spectral function $A(\omega)$ of the attractive polarons (in units of $1/E_n$) versus  $T$ at $k_n a_B = 0.01$ and $k_n a = -1$, computed using the ladder approximation (dashed lines and lighter shading) and our extended model (thick lines and darker shading). For clarity, we added an artificial tiny width of $0.01E_n$ to the spectral lines. The thick solid lines in the plane show the polaron energies, as given by Eq.~\eqref{QPeqn}, and the width of their shading gives the damping $\Gamma$. The dash-dotted lines are the low temperature result $\omega_0(1 \pm \sqrt{Z_0n_\text{ex}/n_0})$. 
}
\end{figure}

The impurity spectral function is  shown in Fig.~\ref{fig:attractivePolaronsKnAM1} as a function of temperature  for $k_na=-1$. We plot results obtained from both the  extended scheme 
and  the ladder approximation. 
For $T=0$, the two schemes give essentially the same result: The spectral function exhibits a single narrow peak corresponding to a quasiparticle, the attractive polaron, with 
energy $\omega\simeq -0.3 E_n$. This is not surprising, since $k_n a_B = 0.01\ll 1$ and very few bosons are excited out of the BEC at $T=0$, so that $\Sigma_1$ is  negligible in both the ladder  and the extended scheme.

Significant differences between the two approximations appear however for $T>0$. 
As shown in Fig.~\ref{fig:attractivePolaronsKnAM1}, while the spectral function in the ladder approximation exhibits a single polaron peak, corresponding to a single quasiparticle solution of 
 $\omega_L$ of \eqref{QPeqn}, the extended  approximation yields \emph{two quasiparticle peaks} in the spectral function. Correspondingly, there are \emph{two} quasiparticle (polaron) solutions to \eqref{QPeqn}, 
 $\omega_\uparrow$ and $\omega_\downarrow$.
Just above $T=0$, the two sharp polaron peaks emerge symmetrically out of the ladder polaron with similar spectral weight. In fact, their residues $Z_\uparrow$ and $Z_\downarrow$ are both close to  $Z_L/2$. Both features can be explained by a  pole expansion valid for $k_na_B\ll1$~\cite{SM}. 
For weak coupling, 
$\omega_{\uparrow,\downarrow} \simeq \omega_0(1 \pm \sqrt{Z_0n_\text{ex}/n_0})$, where $\omega_0=n_0{\mathcal T}_v$ is the energy of a zero momentum impurity dressed by the condensate only, and   $n_{\rm ex}=n-n_0(T)$ is the density of bosons excited out of the condensate~\cite{SM}. 
The energy of the upper polaron $\omega_\uparrow$ increases with $T$, but its spectral weight drops to zero as $T_c$ is approached.
The energy of the lower polaron $\omega_\downarrow$ instead decreases with temperature, reaching a minimum at $T_c$, after which it increases. 
For  $T\ge T_c$ there is no BEC, and the extended and ladder approximation coincide. The width of the lower polaron peak increases with temperature reflecting 
increased damping due to scattering on thermally excited bosons. Eventually, the polaron becomes ill-defined for $T\gg T_c$.

\begin{figure}[t]
	\includegraphics[width=\columnwidth]{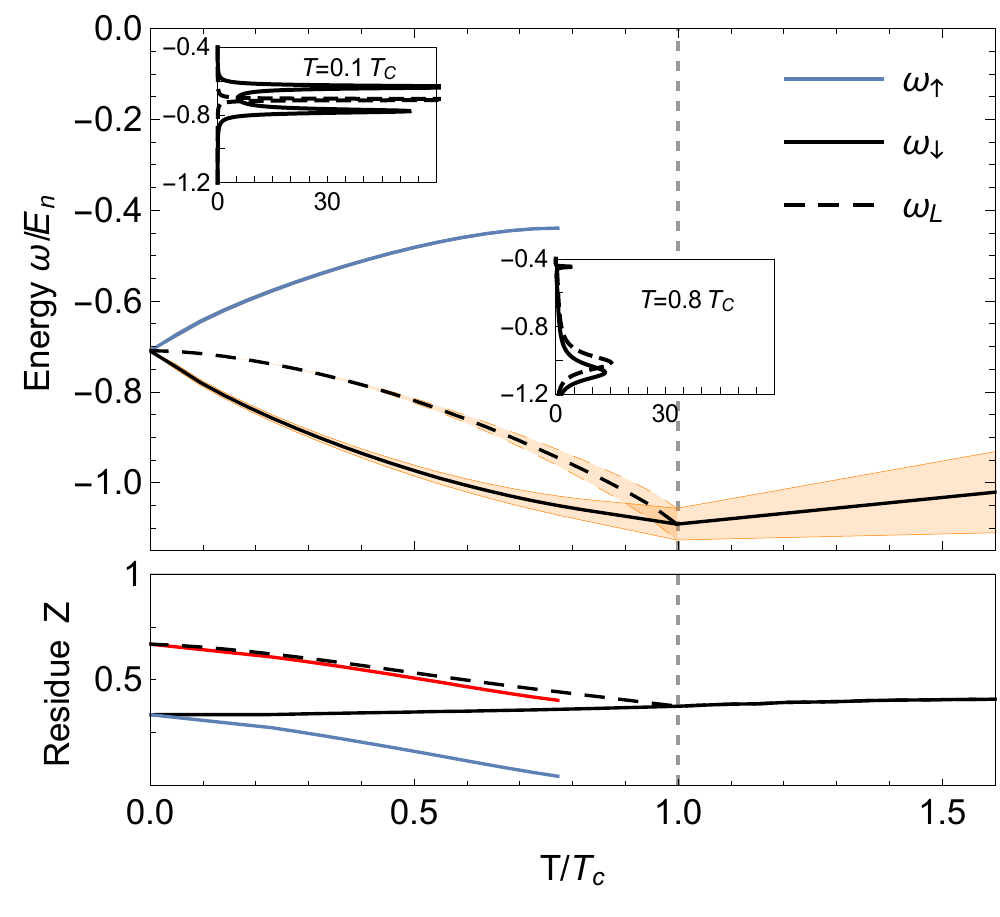}
	\caption{\label{fig:attractivePolaronsUnitarity}
		Energies (top) and residues (bottom) of the attractive polarons for $1/k_n a = 0$ in the ladder approximation (dashed line) and from our extended model (solid lines). The width of the shading around the energies gives the damping $\Gamma$. 
		We stop plotting $\omega_\uparrow$ when its residue is below 5\%. The insets depict the spectral function $A(\omega)$ at $T=0.1T_c$ (left) and $T=0.8T_c$ (right). The red line shows $Z_\uparrow + Z_\downarrow$.
}
\end{figure}
\begin{figure}[h!]
	\includegraphics[width=\columnwidth]{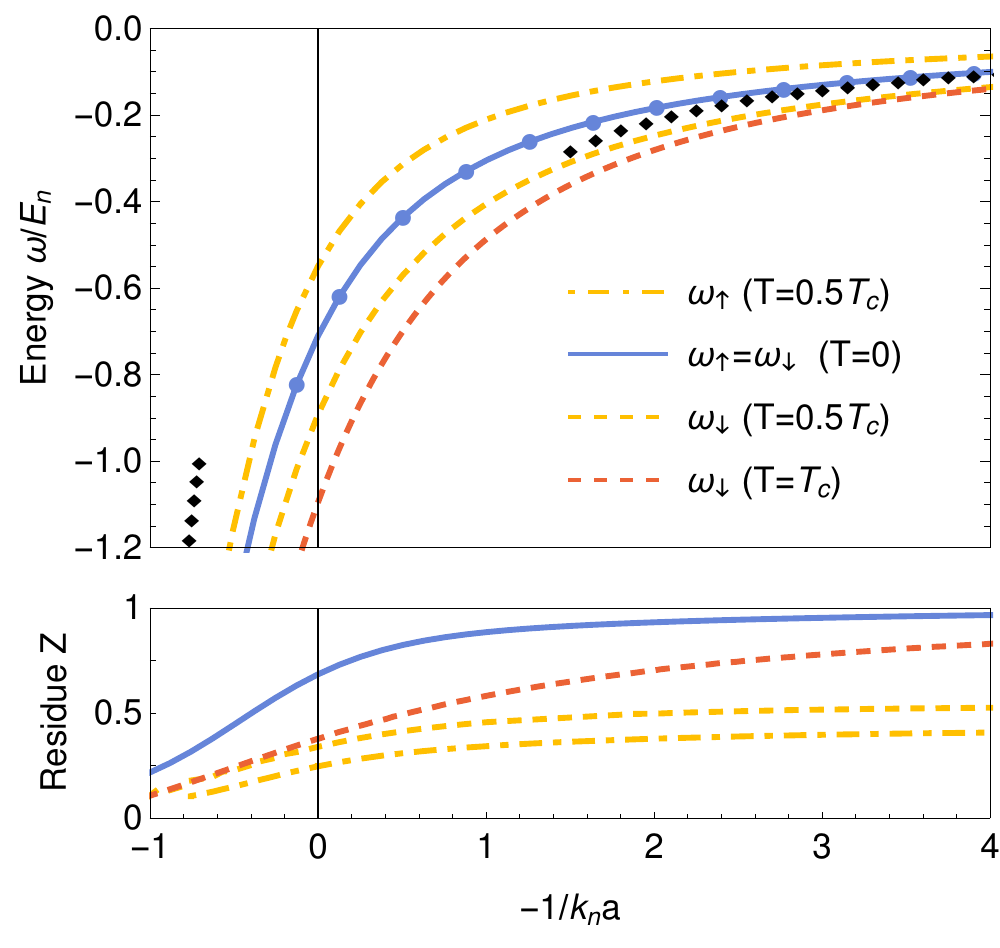}
	\caption{\label{GS_energy_in_the_crossover}
Energy (top) and residue (bottom) of the attractive polarons versus interaction strength, at $T=0$ (blue), $T=0.5T_c$ (yellow), and $T=T_c$ (red). 
The filled blue circles in the energy plot indicate the results of the ladder approximation at $T=0$, and the two black lines (filled diamonds) are the mean field results $\omega=2\pi a/m_r$ (right) and $\omega=-1/(2m_ra^2)+\mu_B$ (left), valid respectively for $-1/k_na\gg1$ and $-1/k_na\ll-1$.}
\end{figure}

\begin{figure}[t]
\includegraphics[width=\columnwidth]{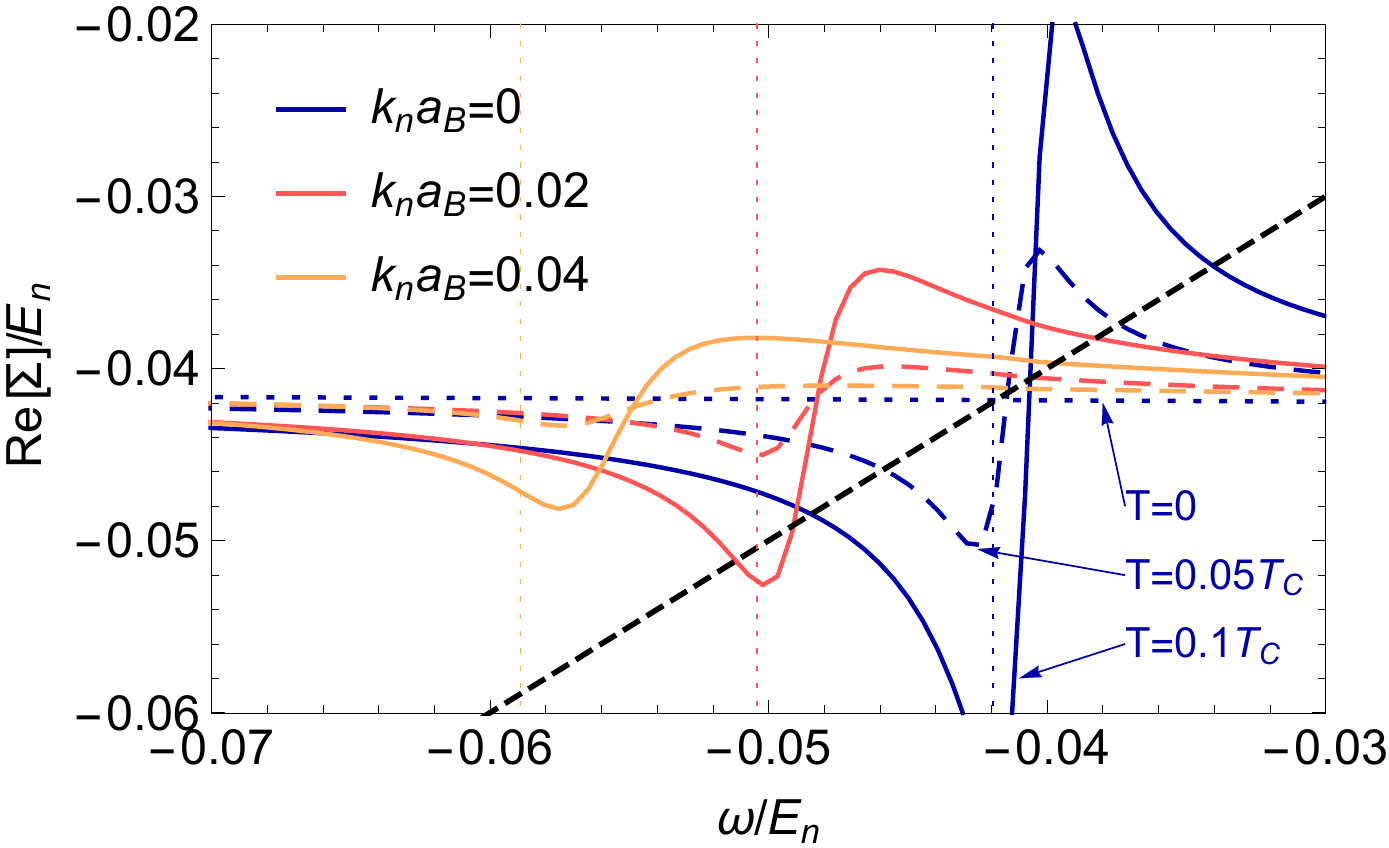}
\caption{\label{fig:self_energy_BCS_limit}
Real part of the self-energy $\Sigma(\bp=0,\omega)$ for  $k_na=-0.1$, at $T=0$ (purple dotted line, almost horizontal), $T=0.05T_c$ (dashed) and $T=0.1T_c$ (solid). The different colors correspond to various values of $k_na_B$. The diagonal black line represents $\omega$, and the vertical lines denote  $n_0(\cT_v-\cT_B)$.
}
\end{figure}

In Fig.~\ref{fig:attractivePolaronsUnitarity}, we show the quasiparticle spectrum and residues at unitarity $1/k_na=0$, obtained by solving \eqref{QPeqn}. This  shows the same physics 
as we found for $k_na=-1$, but scaled to larger binding energies. Using a virial expansion to compute the self-energy for $T\gg T_c$, and assuming also $T \gg 1/m_ra^2$, we obtain
\begin{align}\label{OmegaVirialUnitarity}
\Sigma(0,0)\approx -i \frac{4E_n}{3\pi^{3/2}} \left(\frac{m_B}{m_r}\right)^2\sqrt{\frac{E_n}{T}}.
\end{align}
Thus, at high temperature the polaron energy approaches zero, and the quasiparticle becomes overdamped.
Our numerical results  converge to (\ref{OmegaVirialUnitarity}) for $T\gg T_c$, although such high temperatures are not shown  in Figs.~\ref{fig:attractivePolaronsKnAM1} and \ref{fig:attractivePolaronsUnitarity}. 

In Fig.~\ref{GS_energy_in_the_crossover} we plot the energy  of the attractive polaron as a function of the interaction strength $1/k_na$, for $k_n a_{B} =0.01$ and at various temperatures. 
For all coupling strengths $a \gtrsim a_B $, we observe that the single attractive polaron present at $T=0$ splits into two polarons at intermediate temperatures, of which only the lower one survives at $T\ge T_c$.  

To analyze the origin of the splitting, in Fig.\ \ref{fig:self_energy_BCS_limit} we plot  $\text{Re}[\Sigma(\omega)]$ for $k_na=-0.1$ and various values of $k_na_B$ and $T/T_c$.
Graphically, the quasiparticle solutions (\ref{QPeqn}) are determined by the crossing of $\textrm{Re}[\Sigma(\omega)]$  with the diagonal dashed line giving $\omega$.  
Focus first on the purple lines relevant for a very small value of $k_na_B$. 
For $T=0$ (purple dotted line, almost horizontal) there is only one solution  close to $\omega_0$. 
However, at finite temperatures $\Sigma_1(\omega)$ develops a resonance structure around $\omega_0$ and two further crossings appear. At the middle crossing (the one with $\partial_\omega\rm{Re}[{\Sigma}]>0$) the self-energy has a large imaginary part, so that this does not correspond to a well-defined excitation; as a consequence, the spectrum contains now \emph{two} quasiparticle solutions. 
The resonance at $\omega\sim \omega_0$ arises because the extended matrix $\ctT$ in \eqref{extTmatrix} has a pole at $\wc_p=\epsilon_\bp + \Sigma_0(\wc_p,\bp)$, which is 
very close to $\omega_0$ for $T=0$ and $\bp \rightarrow 0$, since  the number of excited bosons is negligible. For $T>0$, the Bose distribution $f_\bk$ is infrared divergent, 
which in combination with the pole of the $\ctT$ matrix means that the integrand in  \eqref{Sigma1} for $\Sigma_1$ is very large for $\bk \rightarrow 0$,  changing
sign around $\omega\sim\wc\approx\omega_0$.
 This results in the resonance structure shown in Fig.\ \ref{fig:self_energy_BCS_limit}, a completely non-perturbative feature of the extended scheme.
The energy splitting between $\omega_\uparrow$ and $\omega_\downarrow$ increases with temperature, since more particles are thermally excited out of the BEC leading to a more pronounced resonance structure. This explains the behaviour seen in Figs.~\ref{fig:attractivePolaronsKnAM1}-\ref{fig:attractivePolaronsUnitarity}.

The physical interpretation of the resonance feature of $\ctT$ is that the coupling between the impurity and the low-lying Bogoliubov modes gets strongly enhanced by the presence of the macroscopically occupied condensate modes  for $\bk \rightarrow 0$. The coupling to the BEC itself however is via the usual ladder scattering matrix $\cT$. At finite temperatures
 $0<T<T_c$, the impurity couples to both the BEC and a macroscopic amount of excited bosons, and the hybridization between these two terms leads to the emergence of two quasiparticles.

Increasing $k_na_B$ has two main effects on the resonance structure of $\Sigma$. 
First, it becomes less pronounced, since the low-energy density-of-states of the BEC decreases due to the stiffening of the Bogoliubov dispersion. 
Second, it moves to lower energies because of the increased effect of the coherence factors $u_\bk,v_\bk$.
Indeed, in the weak coupling limit the resonance is approximately located at $n_0(\mathcal{T}_v-\mathcal{T}_B)$~\cite{SM} 
(these energies are plotted as vertical dashed lines in Fig.\ \ref{fig:self_energy_BCS_limit}). 
It follows from these two effects that at low temperatures only \emph{one} quasiparticle solution (the upper polaron) survives in the limit of weak impurity-bath 
coupling $|a| \lesssim a_B\ll 1/k_n$, and the energy of this solution approaches the first order result $n\mathcal{T}_v$, in agreement with perturbation theory~\cite{SM}.

An important question concerns whether three-body losses wash out the effects predicted in this paper. Three-body losses were investigated in the Aarhus and JILA experiments. Both groups found that they have surprisingly small effects, in the sense that the observed spectra could be explained theoretically without introducing three-body losses. Since our predicted temperature shifts and splittings have the same order of magnitude as the polaron energy at $T=0$, we conclude that  their observation is likely robust towards three-body losses.

\paragraph{Conclusions and outlook.--}
Using a  diagrammatic resummation scheme designed to include scattering terms that are crucial at finite temperature, we showed that the Bose polaron has a highly 
non-trivial temperature dependence.
For attractive interactions, the polaron splits into two quasiparticle states  for $0<T<T_c$.
The energy of the lower polaron displays a minimum at $T_c$, whereas the energy of the upper one increases with $T$ until its residue vanishes at $T_c$. 
These effects arise due to the coupling of the impurity to the Bogoliubov spectrum, whose population has an infrared divergence for finite temperature, and the enhancement of the coupling due to the presence of the condensate mode. 
Both the splitting of the polaron into two quasiparticles and the temperature dependence of their energies should be  observable in  experiments with ultracold gases, provided that specific care is taken to minimize the frequency and trap averaging of the spectral signal~\cite{Jorgensen2016,Hu2016}. 

Here we considered only negative energies.
It would be interesting to investigate whether the discrepancy found in Ref.~\cite{Jorgensen2016} between theory and experiment at positive energies could be due to finite temperature effects.
A complete treatment of this subtle point requires a self-consistent calculation, which is beyond the scope of the present work, but a brief discussion of the matter is presented in the Supplemental Material~\cite{SM}.

Let us conclude by noting that the physical mechanism leading to the two main results of our paper, the strong temperature dependence and the quasiparticle splitting, is generic: both effects are caused by the condensed Bose gas breaking a continuous symmetry below $T_c$, so that a gapless mode appears, leading to a dramatic change in the low energy density of states to which the impurity couples.
Our findings are therefore relevant to a wide class of systems consisting of an impurity immersed in a medium breaking a continuous symmetry. This includes impurities in liquid Helium~\cite{BaymPethick1991book}, normal and high-$T_c$ superconductors~\cite{Dagotto1994}, ultracold Fermi superfluids~\cite{Wei2015}, 
and quantum magnets~\cite{Fiete2003}. A similar splitting of a fermionic quasiparticle into two modes due to the coupling to a linear bosonic spectrum has indeed 
been predicted in hot quark-gluon plasmas, and in Yukawa and QED theories~\cite{Klimov1982,Weldon1982,Weldon1989,Baym1992}. 
Due to its collective nature, the emergent quasiparticle  was dubbed {\it plasmino}~\cite{Braaten1992}. Our results therefore provide an avenue to study this interesting
 prediction in the controlled environment of a quantum gas.

\vspace{5mm}
\begin{acknowledgements}
We wish to thank J. Arlt, G. Baym, A. Fetter, J. Levinsen, M. Parish,  C. Pethick, P. Pieri, R. Schmidt, and L. Tarruell for very insightful discussions.
This work has been supported by 
Spanish MINECO (Severo Ochoa SEV-2015-0522, FisicaTeAMO FIS2016-79508-P), 
the Generalitat de Catalunya (SGR 874 and CERCA),
Fundaci\'o Privada Cellex, 
and EU grants EQuaM (FP7/2007-2013 323714), 
OSYRIS (ERC AdG), 
QUIC (H2020-FETProAct-2014 641122), 
and SIQS (FP7-ICT-2011-9 600645).
NG is supported by a ``la Caixa-Severo Ochoa" PhD fellowship.
PM acknowledges funding from the ``Ram\'on y Cajal" program and the Simons Foundation, and the kind hospitality of the Aspen Center for Physics, where part of this work was realized, and which is supported by NSF grant PHY-1607611. GMB acknowledges the support of the Villum Foundation via Grant No. VKR023163.
\end{acknowledgements}

\bibliography{Polarons} 

\begin{thebibliography}{52}
\bibitem [{foo()}]{footnote}%
  \BibitemOpen
  \href@noop {} {}\bibinfo {note} {Here $2j$ is an even (odd) integer number
  when the impurity is bosonic (fermionic). However, since we consider a single
  impurity, it does not matter whether it is fermionic or bosonic.}\BibitemShut
  {Stop}%
\end{thebibliography}

\end{document}